\documentclass[%
 aip,
 sd,%
 amsmath,amssymb,
 reprint,%
]{revtex4-1}

\usepackage{graphicx}
\usepackage{dcolumn}
\usepackage{bm}

\usepackage{amsmath}
\usepackage{cases}
\usepackage{amsbsy}
\usepackage{amssymb}
\usepackage{amscd}
\usepackage{amsfonts}
\usepackage{graphics}
\usepackage{subfigure}
\usepackage{epsfig}
\usepackage{float}
\usepackage{tabularx}
\usepackage{ctable}
\usepackage{fancyhdr}
\usepackage{pstricks}

\usepackage[overload]{empheq}
\usepackage{cleveref}

\usepackage{natbib}
\usepackage{graphicx}
\usepackage{dcolumn}
\usepackage{bm}
\usepackage[mathlines]{lineno}

\def	\D	{{\mathcal{D}}}
\def	\R	{{\mathcal{R}}}

\def	\Vm		{{\langle V(t) \rangle}}

\graphicspath {{./Figures/}}

\begin{document}

\preprint{AIP/123-QED}

\title[Thermo-Electric Nonlinear Diffusion]{Nonlinear diffusion \& thermo-electric coupling in a  two-variable model of cardiac action potential\\}

\author{A. Gizzi}
\affiliation{Department of Engineering, University Campus Bio-Medico of Rome, Unit of Nonlinear Physics and Mathematical Modeling, Via A. del Portillo 21, 00128 Rome, Italy}

\author{A. Loppini}
\affiliation{Department of Engineering, University Campus Bio-Medico of Rome, Unit of Nonlinear Physics and Mathematical Modeling, Via A. del Portillo 21, 00128 Rome, Italy}

\author{R. Ruiz-Baier}
\affiliation{Mathematical Institute, University of Oxford, A. Wiles Building, Woodstock Road,  OX2 6GG Oxford, UK}

\author{A. Ippolito}
\author{A. Camassa}
\author{A. La Camera}
\author{E. Emmi}
\author{L. Di Perna}
\author{V. Garofalo}
\affiliation{Department of Engineering, University Campus Bio-Medico of Rome, Unit of Nonlinear Physics and Mathematical Modeling, Via A. del Portillo 21, 00128 Rome, Italy}


\author{C. Cherubini}
\affiliation{Department of Engineering, University Campus Bio-Medico of Rome, Unit of Nonlinear Physics and Mathematical Modeling, Via A. del Portillo 21, 00128 Rome, Italy}
\affiliation{International Center for Relativistic Astrophysics (I.C.R.A.), University Campus Bio-Medico of Rome, Italy}


\author{S. Filippi}
\affiliation{Department of Engineering, University Campus Bio-Medico of Rome, Unit of Nonlinear Physics and Mathematical Modeling, Via A. del Portillo 21, 00128 Rome, Italy}
\affiliation{International Center for Relativistic Astrophysics (I.C.R.A.), University Campus Bio-Medico of Rome, Italy}

\date{\today}

\begin{abstract}
This work reports the results of the theoretical investigation of nonlinear dynamics and spiral wave breakup in a generalized two-variable model of cardiac action potential accounting for thermo-electric coupling  and diffusion nonlinearities. As customary in excitable media, the common $Q_{10}$ and Moore factors are used to describe thermo-electric feedback in a 10-degrees range. Motivated by the porous nature of the cardiac tissue, in this study we also propose a nonlinear Fickian flux formulated by Taylor expanding the voltage dependent diffusion coefficient up to quadratic terms. A fine tuning of the diffusive parameters is performed a priori to match the conduction velocity of the equivalent cable model.
The resulting combined effects are then studied by numerically simulating different stimulation protocols on a one-dimensional cable. Model features are compared in terms of action potential morphology, restitution curves, frequency spectra and spatio-temporal phase differences. Two-dimensional long-run simulations are finally performed to characterize spiral breakup during sustained fibrillation at different thermal states.
Temperature and nonlinear diffusion effects are found to impact the repolarization phase of the action potential wave with non-monotone patterns and to increase the propensity of arrhythmogenesis.
\end{abstract}

\maketitle

\begin{quotation}
Systemic temperature is kept approximately constant through many delicate physiological feedbacks. In the heart, temperature changes greatly affect the features of the excitation wave and occur because of pathological conditions, during surgery or because of various unfortunate events. It is therefore practically relevant to determine how thermo-electric feedbacks can alter normal cardiac pacing or sustain fibrillating scenarios. Recent studies have demonstrated that structural heterogeneity of cardiac tissue is a key ingredient for understanding dispersion of repolarization and tendency to arrhythmogenesis. Here we investigate whether a generalized two-variable reaction-diffusion model of cardiac electrophysiology can highlight the dual nonlinear effects induced by thermo-electricity and diffusive nonlinearities under several simulated tests. Our aim is to provide evidence of non negligible differences in the spatio-temporal behavior of the system when these contributions are considered and to stimulate the investigation of more reliable physiological cardiac models.
\end{quotation}

\section{\label{sec:introduction}Introduction}

Understanding how transition from normal rhythm to fibrillation occurs within the heart has been for many years a primary focus of research in cardiac electrophysiology, theoretical physics, mathematical modeling as well as clinical practice~\cite{winfree:1987,karma:1991,karma:1992,karma:1993,karma:1994,karma:1994D,fenton:1998,pumir:1998,fenton:2002,fenton:2005}. 
Cardiac modeling, in fact, experienced a fast-paced growth of sophisticated mathematical models~\cite{fenton:2008} that can accurately reproduce several electrophysiological features observed in isolated cardiomyocytes, small tissues as well as whole organs both in healthy and pathological conditions~\cite{fenton:1998,fenton:2005,luther:2011,saez:2016}. 
However, several open questions still need to be answered~\cite{clayton:2011}, e.g.: what type of electrical activity induce spiral onset and provoke spiral breakup?
Experimental evidences of the complex spatio-temporal alternans dynamics arising during fast pacing and their relation with tissue heterogeneity~\cite{pastore:1999, pastore:2000,gizzi:2013}, in particular, call for generalized theoretical approaches and mathematical modeling tools~\cite{chen:2017} able to provide a mechanistic description of these phenomena.

Virtually all mathematical models of excitable biological media derive from the seminal work of Hodgkin and Huxley on the squid giant axon and are typically formulated in terms of local kinetics and diffusive spatial features, either in their monodomain or bidomain version~\cite{bendahmane:2006,potse:2006,keener,pullan,ruiz:2010}. The resulting set of equations constitute a nonlinear reaction-diffusion (RD) system in which spatial propagation of electrical potentials are coupled with evolution equations describing the ion-channel, exchanger and pump gate dynamics~\cite{niederer:2011}.
On these bases, spatio-temporal irregularities in excitable media have been studied from several points of view~\cite{vinet:1990,biktashev:1999} and multiple approaches have been proposed to understand and control spiral waves dynamics in heterogeneous active media~\cite{pumir:2004,pumir:2005,pumir:2010,biktasheva:2000,dupraz:2014,quail:2014,biktasheva:2015}.

Cardiac tissue, as active biological medium, is inherently characterized by multiphysics couplings and a multiscale structure~\cite{qu:2014}. Cardiomyocytes, in fact, communicate via voltage-sensitive gap junctions proteins~\cite{dehin:2006} affecting the microscopic propagation timing~\cite{rohr:2004} and inducing serious cardiac diseases when genetically modified~\cite{severs:2008}.
In addition, local heterogeneities and anisotropies play a major role in action potential spread and propagation~\cite{potse:2007}. Temperature, in particular (as primary theme of the present work), affects the time constant of the local kinetic reactions thus inducing an enhanced level of heterogeneity to the tissue~\cite{cherry:2007}. 
For these reasons, we focus here on a well established thermo-electric coupling known for greatly affecting spatio-temporal dynamics of excitable biological tissues~\cite{gizzi:2010,fenton:2013,filippi:2014} and for being of primary importance in cardiology~\cite{bigelow:1950,reuler:1978,lancet:1997}. 
%
%
Other approaches have been recently proposed in order to incorporate local spatial features within a computational framework, e.g.  fractional diffusion operators~\cite{bueno-orovio:2014,bueno-orovio:2014a,bueno-orovio:2015,bueno-orovio:2016} or statistical mechanics of cell-cell interaction~\cite{deshpande:2006,deshpande:2008}. 

In healthy cardiac tissue dispersion usually has a weak role and the pulse speed is insensitive to the post-repolarization state due to a previous pulse. However, this condition does not hold when the tissue is subject to fast pacing and the diastolic interval (the time laps between the end of pulse and the beginning of the next pulse) is short.

As theoretical study, in this work we deliberately assume a minimal  formulation comprehensive of generalized multiphysics couplings. In detail, we make use of the two-variable phenomenological Karma model~\cite{karma:1993} thus incorporating a Taylor expanded diffusion function up to quadratic terms and introducing the classical $Q_{10}$ and Moore factors describing the thermo-electric coupling~\cite{fenton:2013,filippi:2014}. 
The object of the numerical analysis is to investigate the emerging spatio-temporal features of linear versus nonlinear diffusion within a thermo-electric framework. We will demonstrate that the combined nonlinear effects modify the spatio-temporal behaviors of the system without impacting its local dynamics. 

From the numerical point of view, we provide a fine tuning of the diffusive parameters with respect to a fixed conduction velocity and we show the ability of our formulation to avoid numerical oscillations as well as artificial discontinuities present in the classical porous medium approach~\cite{hurtado:2016}. We analyze a number of electrophysiological features associated with the action potential (AP) wave thus comparing linear versus nonlinear diffusion (LD, NLD) models at different thermal states: i) action potential duration (APD) and ii) conduction velocity (CV) restitution curves, iii) time and space phase differences (PD) and iv) dynamic averages of the membrane voltage ($\Vm$).  One-dimensional (1D) cable, ring and two-dimensional (2D) computational domains are considered in varying domain sizes, boundary conditions and local distribution of the temperature parameter $T$.


%

\section{Methods}
\label{sec:methods}

\subsection{Generalized RD Model}
\label{sec:RD}
We present the generalized formulation of a two-variable RD cardiac model

\begin{subequations}
\begin{align}[left ={\empheqlbrace}] 
	\label{eq:genKarmaV}
	\dfrac{\partial V}{\partial t} &= 
	\nabla \cdot D(V) \, \nabla V + \dfrac{f(V,n)}{\tau_V(T)} + I_{\rm ext}
	\\
	\label{eq:genKarmaN}
	\dfrac{dn}{dt} &= \dfrac{g(V,n)}{\tau_n (T)}
\end{align}
\end{subequations}

\noindent
where the variable $V$ is a dimensionless representation of the transmembrane voltage and the variable $n$ plays the role of a slow recovery current. Here $\nabla$ and $\nabla\cdot$ denote the gradient and divergence operators, respectively. According to Karma~\cite{karma:1993,karma:1994}, the nonlinear reaction functions $f(V,n),g(V,n)$ identify a generalized FitzHugh-Nagumo model qualitatively that reproduce generic restitution and dispersion properties of cardiac tissue: 

\begin{subequations}\label{eq:reactions}
\begin{align}[left ={\empheqlbrace}] 
	f(V,n) &= -V + [V^* - \D(n)] h(V)
	\\
	g(V,n) &= \R(n) \theta(V-V_n) - [1-\theta(V-Vn)] n
	\\
	h(V) &= [1 - \tanh(V-V_h)] \dfrac{V^2}{2}
	\\
	\label{eq:RnDn}
	\R(n) &= \dfrac{1-(1-e^{-Re}) n}{1-e^{-Re}}
	,\qquad\quad
	\D(n) = n^M
\end{align}
\end{subequations}

\noindent
where $\theta(x)$ denotes the standard Heaviside step function (i.e. $\theta(x)=0$ for $x\le0$ and $\theta(x)=1$ for $x>0$), the nondimensional membrane voltage $V$ varies in $[0\div4]$ and the rest state of the system corresponds to $(V,n)=(0,0)$. 
The restitution function, $\R(n)$, controls the relationship between two time intervals: the time-frame between the end of an action potential pulse and the beginning of the next one (diastolic interval, DI) and the duration of the next action potential pulse (APD). The dispersion function, $\D(n)$, prescribes the relation between the instantaneous speed of the AP front-end at a given spacial point and the time elapsed since the back-end of a previous pulse passed the same location.
Provided the specific functions \eqref{eq:RnDn}, the restitution is controlled by the parameter $Re$ while the dispersion is controlled by the parameter $M$.

We account for a self diffusion quadratic law, $D(V)$, and thermal coupling, $\tau_V(T),\tau_n(T)$, via the following constitutive relations:

\begin{subequations}
\begin{align}[left ={\empheqlbrace}] 
	D(V) &= D_0 + D_1 V + D_2 V^2
	\label{eq:DV}
	\\
	\label{eq:tauT}
	\tau_V (T) &= \dfrac{\tau_V^0}{1 + \beta(T-T_0)}
	,\quad
	\tau_n (T) = \dfrac{\tau_n^0}{Q_{10}^{(T-T_0)/10}}
\end{align}
\end{subequations}

\noindent
Such an approach generalizes the porous medium formulation~\cite{hurtado:2016} both in terms of RD properties and thermo-electric effects~\cite{fenton:2013} and extends the restitution and dispersion properties of the original model.

In detail, the functional form~\eqref{eq:DV} accounts for a basic constant diffusion $D_0$ enriched by linear, $D_1$, and quadratic, $D_2$, self diffusion terms that contribute to the diffusive flux according to the voltage field level $V$. When the voltage is low (close to the resting state) the additional contributions are negligible and the equation boils down to the standard cable equation with constant diffusion $D_0$. However, when the voltage is at the plateau phase (depolarized state) then these two additional contributions affect the diffusive flux and consequentially speed and wavelength of the propagating wave. 
This particular feedback, well known in the context of animal dispersal and population growth~\cite{murray,cherubini:2012}, is here introduced to mimic, in a phenomenological way, the nonlinear dependence of gap junctions conductance on the transjunctional voltage occurring at the sub-micron scale~\cite{rohr:2004}. It is worth noticing here that we do not aim at reproducing the exact physical nature of gap junction couplings at the macroscopic level (whose description would require a more complex multiscale homogenization procedure), but we investigate the properties of our generalized formulation by analyzing its nonlinear behaviors via the classical tools used for dynamical systems.

On a similar phenomenological ground, well established is the nonlinear coupling adopted for the thermo-electric model~\eqref{eq:tauT}. In particular, an Arrhenius exponential law, $Q_{10}$, is introduced into the dynamics of the gating variable $\tau_n(T)$~\eqref{eq:tauT}$_2$ whereas a linear deviation (referred to as Moore term) is considered for the time constant $\tau_V(T)$ of the membrane voltage~\eqref{eq:tauT}$_1$. These additional thermal factors contribute to adapt the restitution and dispersion time scales according to the selected temperatures. They greatly affect the duration of the AP wave as well as its conduction velocity~\cite{fenton:2013}. Therefore, the concomitant effect of nonlinear diffusion and thermo-electric coupling will result in enhanced properties for the dynamical systems that we aim at exploring in the present contribution.


\subsection{Parameter's Space Analysis}

In order to avoid a wide and unstructured range of free parameters, we set the thermal coefficients according to the experimental fitting reported in Fenton et al.~\cite{fenton:2013} and relative to endocardial optical mapping recordings~\cite{gizzi:2013}.
Accordingly, we study the features of our thermo-electric dynamical system for a fixed set of model parameters such that extended comparisons are performed between linear and nonlinear diffusion operators only at different thermal states.
To this aim, a preliminary fine tuning of the nonlinear diffusion parameters $D_0,D_1,D_2$ was conducted at physiological temperature, i.e. $T=37^\circ$ C, in order to reproduce the original maximum conduction velocity of $30~cm/s$ computed from the linear model with constant diffusion equal to $\tilde{D}=1.1~cm^2/s$. We remark that for $T=37^\circ$C the thermo-electric coupling has no effects (both $Q_{10}$ and Moore terms are equal to 1).

As preliminary tuning, several combinations of the three parameters were tested by adopting as ultimate limiting condition the order of magnitude of the three coefficients, i.e. $D_0\gg D_1>D_2$. We remark that, for the nondimensional phenomenological model at hand, the physical dimensions of these coefficients are homogeneous, i.e. $[cm^2/s]$.
In the calculation we started using the value $D_0=\tilde{D}$ from the original model, then we introduced one coefficient at a time (maintaining the other at 0) and lowering the value of $D_0$. Finally, we analyzed the three parameters together and we chose the optimum combination respecting the expected orders of magnitude.
Such a fitting strategy, though not unique, is plausible since we aimed at introducing small deviations with respect to the linear model (cable equation) in a minimal simplified setting.
The complete list of model parameters is provided in Tab.~\ref{tab:params}.

\subsection{Restitution Protocols}
The numerical analysis of the 1D model was conducted via two alternative activation protocols, i.e.: pacing down (or full restitution) and S1S2. The two protocols were used for computing the APD and CV restitution curves and for extracting phase differences between linear and nonlinear diffusion cases.

The full restitution protocol consisted of periodic electrical activations induced through the square wave current $I_{\rm ext}$ at the left boundary of the cable with a constant amplitude equal to 2, a constant duration of $2~ms$ but varying its duty cycle. For a fixed pacing cycle length (CL), $n= 10$ stimulations were delivered before CL was decreased. In particular, the protocol was applied for each selected temperature and for LD and NLD models. 
In order to compute the fine transition of the CV-CL restitution, the applied protocol consisted in the sequence:
CL$=800\div-2\div100~ms$,
a cable of length $L=15~cm$ was adopted in order to ensure the complete evolution of the propagating wave.
The protocol ended when capture of activation was lost and was implemented via finite differences Fortran routines with an explicit time integration scheme of first order, a fixed $dt=0.1~ms$, and fixed $dx=0.025~cm$ (equivalent results were obtained with finite element implementations using piecewise linear approximations for every field and a semi-implicit discretization in time).

The S1S2 protocol consisted of the series of a constant stimulation at CL$_{\rm S1}=450~ms$ followed by a single stimulation spaced with a shorter period CL$_{\rm S2}<$CL$_{\rm S1}$. 
The sequence of CL$_{\rm S1}$ was: 
$450\div-50\div350~ms$,
$350\div-10\div200~ms$.
Also for this case, since the spatial distribution of AP is necessary for the analysis, a  cable of length $L=10~cm$ was considered. In this case, the minimum CL ensuring correct pacing was $280~ms$, higher than those reached during the full restitution. This protocol was implemented in finite elements with mesh size $dx=0.025~cm$ and maximum time step $dt=0.1~ms$.

\subsection{Averaged Action Potential and Gradient Operator}
In the following analysis we make use of two additional quantities related to the membrane voltage $V$ and to the recovery variable $n$. We considered the spatial average of the membrane voltage, $\langle V(t) \rangle$, computed as function of time on both 1D and 2D domains. Accordingly, the time course of the average signal was investigated by means of the FFT transform such that its spectrum was studied.
In addition, the sharp gradients of the recovery variable $n$ were computed for 2D simulations in order to identify the numerosity of localized regions with high gradients (spiral cores) during long run simulations thus to compare LD and NLD model spatio-temporal behaviors.

\begin{table}[htp]
\centering 
\caption{Model parameters.} 
\begin{tabular}{lcrllcr}
\hline\noalign{\smallskip} 
$V_h$ & 3 & &&\quad
$V_n$ & 1
\\
$V^*$ & 1.5415 & &&\quad
$Q_{10}$ & 1.5
\\
$\tau_V$ & 2.5 & [ms] &&\quad
$\beta$ & 0.008
\\
$\tau_n$ & 250 & [ms] &&\quad
$D_0$ & 0.85 & [cm$^2$/s]
\\
$Re$ & 0.9 &  &&\quad
$D_1$ & 0.09 & [cm$^2$/s]
\\
$M$ & 9 & &&\quad
$D_2$ & 0.01 & [cm$^2$/s]
\\
\hline\noalign{\smallskip}
\end{tabular}
\label{tab:params}
\end{table}

\begin{figure*}[t!]
	\subfigure[]{\includegraphics[width=0.4\textwidth]{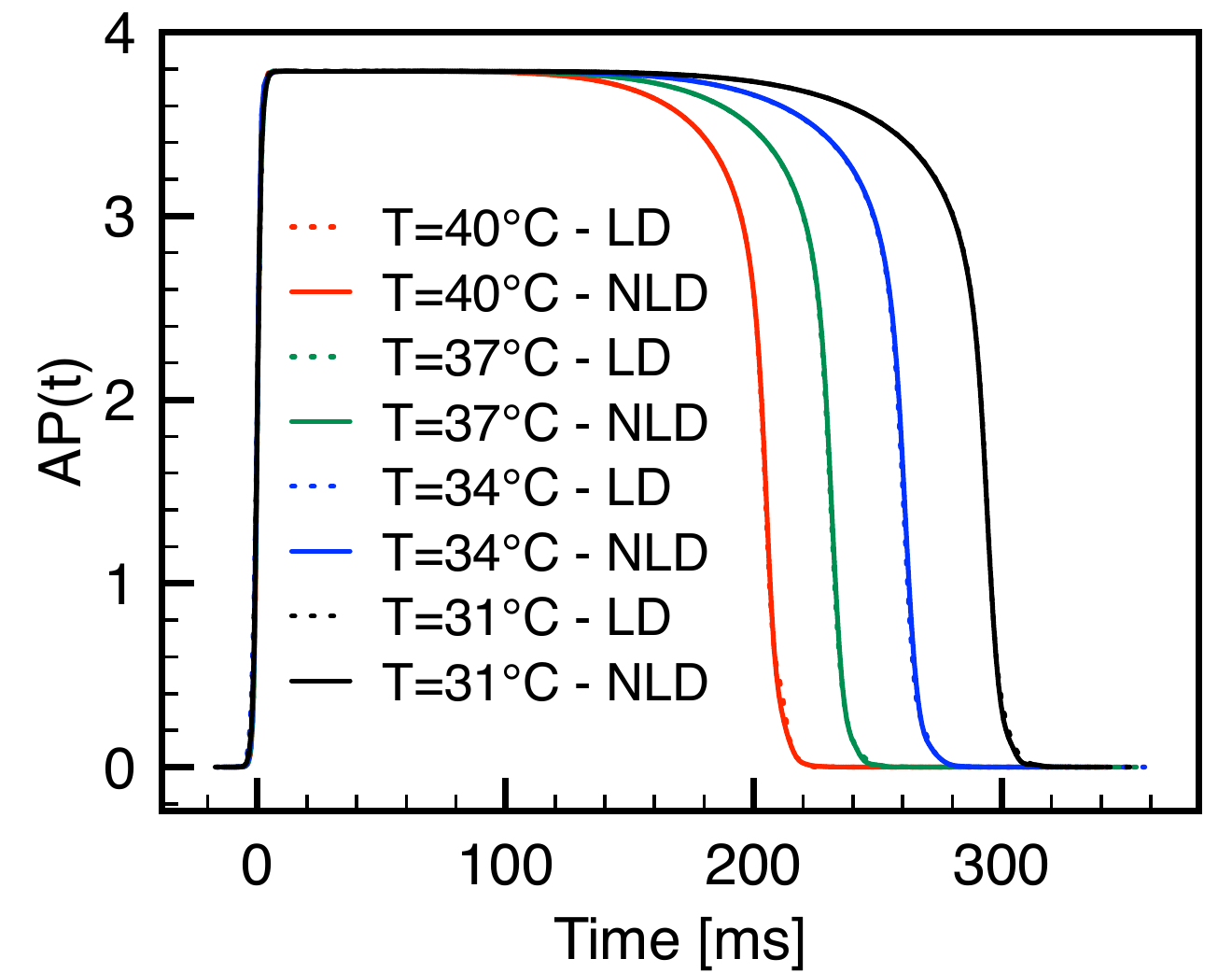}\label{fig:cableT}}
	\subfigure[]{\includegraphics[width=0.4\textwidth]{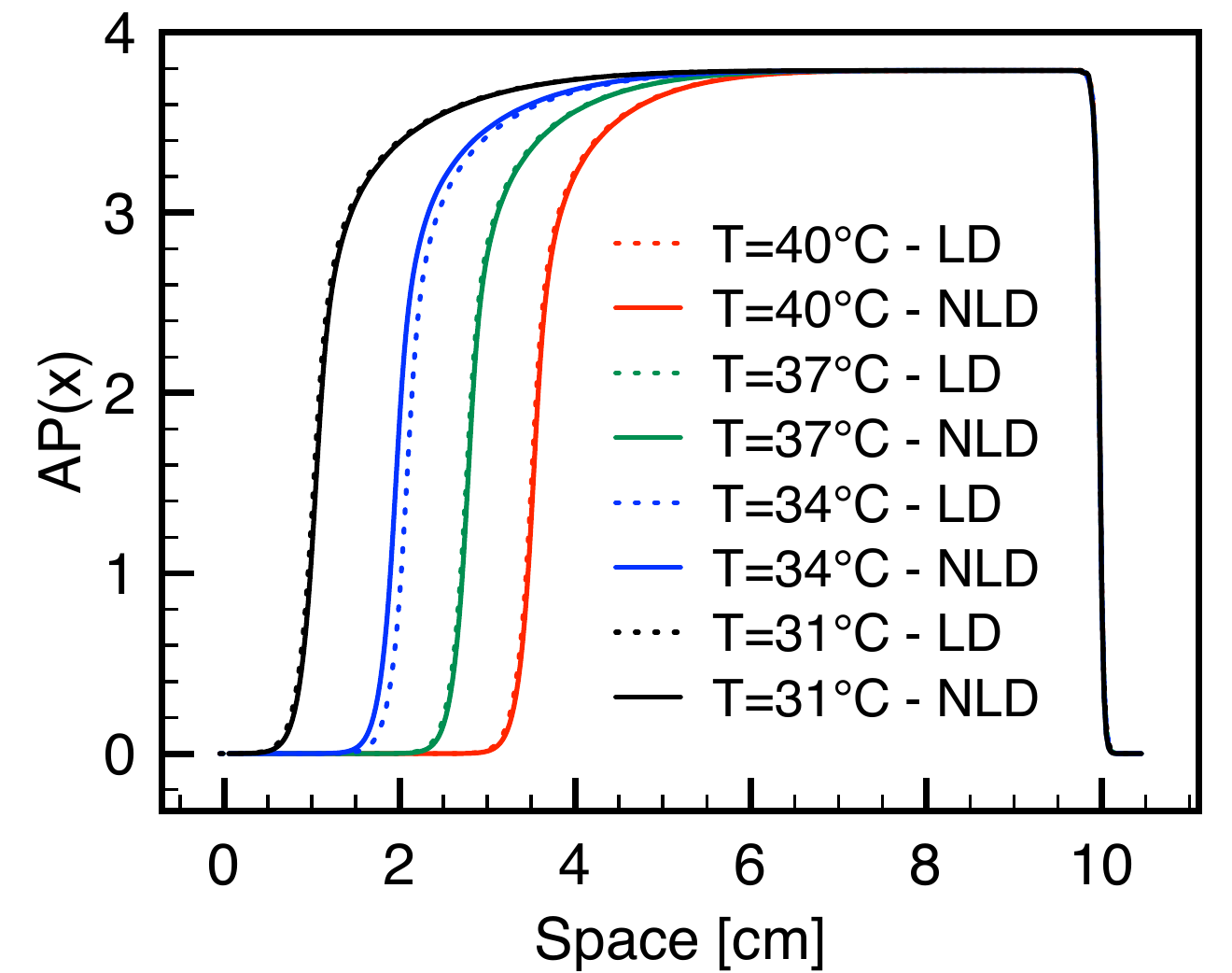}\label{fig:cableS}}
	\\
	\subfigure[]{\includegraphics[height=0.24\textwidth]{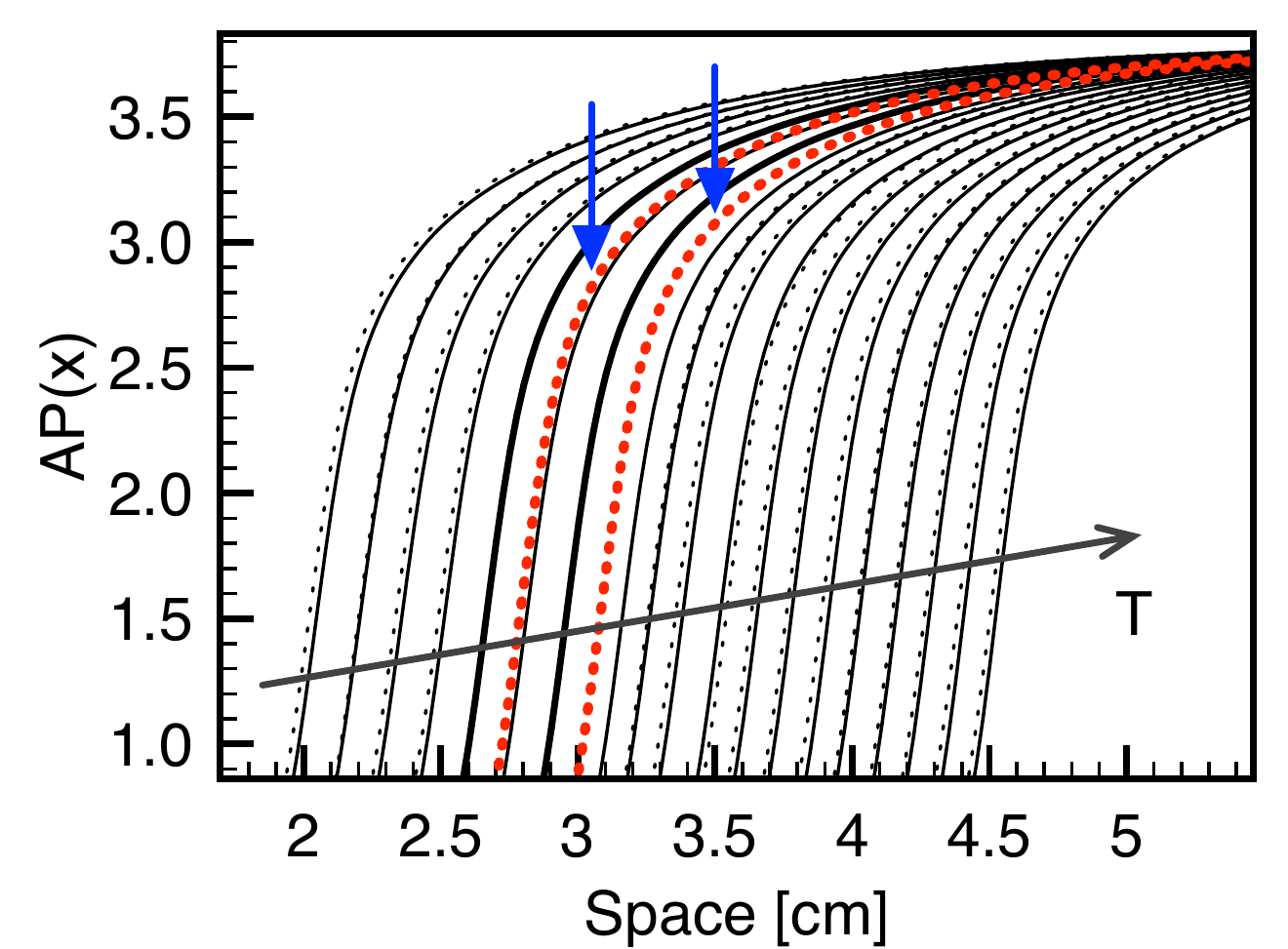}\label{fig:1c}}
	\subfigure[]{\includegraphics[height=0.24\textwidth]{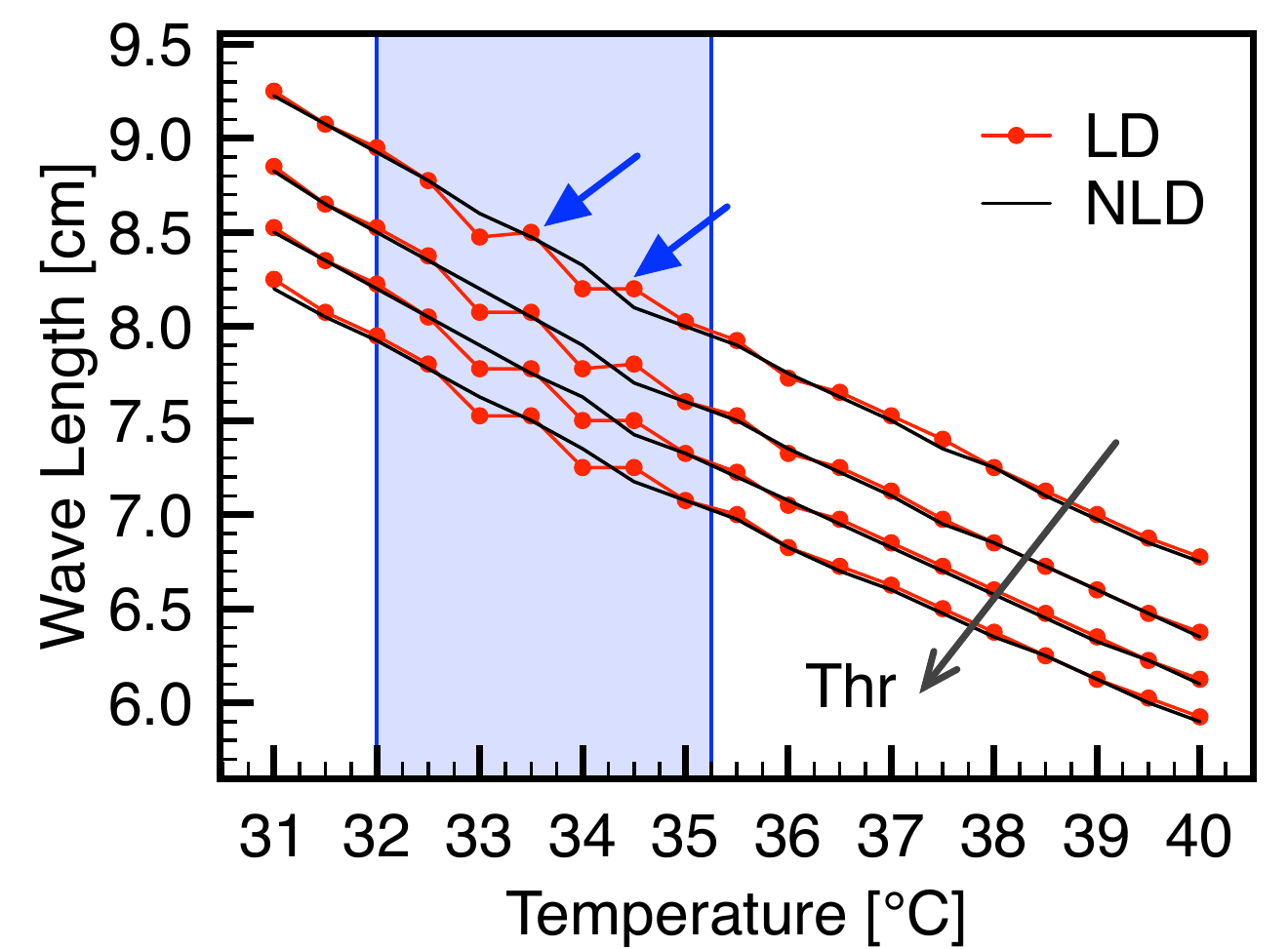}\label{fig:1d}}
	\subfigure[]{\includegraphics[height=0.24\textwidth]{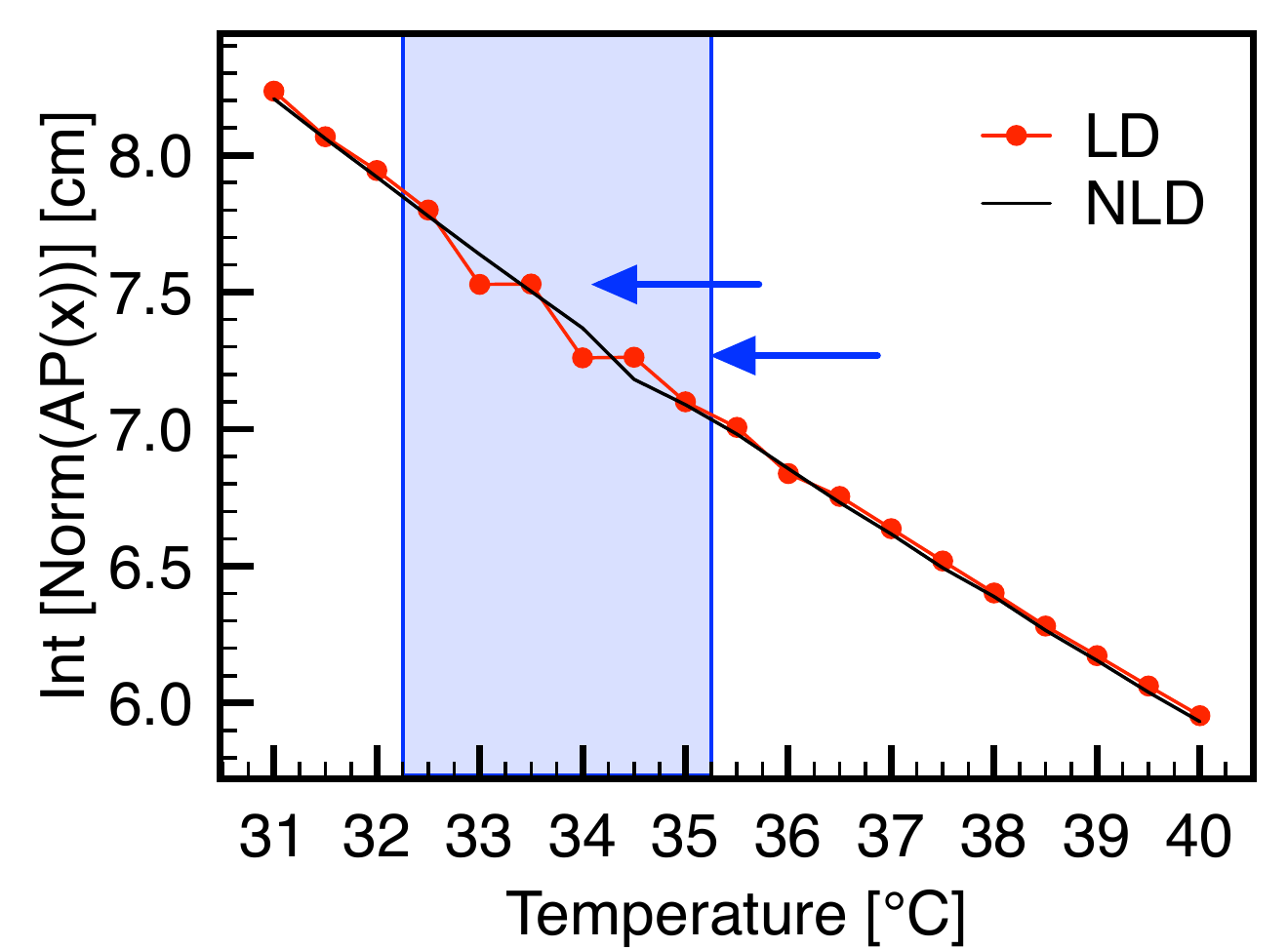}\label{fig:1e}}	
	\caption{\label{fig:cable} 
	Finite element numerical simulations of a 1D cable model of length $L=25~cm$ stimulated on the left boundary via an external squared current $I_{\rm ext}$ ($\Delta t=2~ms$, amplitude 2). Space and time snapshots are taken when the wave is fully developed and travels at constant velocity (mesh size $dx=0.03~cm$ and maximum time step $dt=0.01~ms$ ensure stable conduction velocity).
	Comparison of the AP wave morphology in time (a) and space (b) at selected temperatures. Linear (dashed) and nonlinear (solid) diffusion model are compared.
	 Wave overlapping is obtained on the basis of the wave front threshold AP$=1$ (axes rescale is provided for comparative purposes). 
	 (c) Zoomed view of the back-end space morphology of the excitation wave (LD-dashed, NLD-solid) for a fine gradient of temperatures (step of $0.5^\circ$C).
	 (d) Wave length calculation based on panel (c) for different threshold (Thr) levels.
	 (e) Line integral of the normalized AP spatial distribution based on panel (c).
	 Blue shadowed region highlight the range of temperatures in which the highest differences are observed between LD and NLD model.
	 }
\end{figure*}

\section{\label{sec:results}Results}
We start by comparing the one-dimensional spatio-temporal features of the AP wave arising from the nonlinear thermo-electric model in cable (Neumann zero flux) and ring (periodic boundary conditions) domains. We conclude with 2D long run simulations of spiral breakup dynamics. An extended analysis is conducted comparing linear (LD) and nonlinear (NLD) diffusion at selected temperatures based on the indicators described in the previous section and enriched by the temporal and spatial analysis of AP wave phase differences.
In the description of the results we will make use of a unique color code associated with the temperature parameter: $T=40,37,34,31^\circ$ C will be shown in \emph{red, green, blue, black}, respectively. In addition, when not specified, LD and NLD quantities will be shown as \emph{dashed} and \emph{solid line}, respectively.

\subsection{\label{sec:1Dcable}1D Cable Analysis}

Figures~\ref{fig:cableT} and \ref{fig:cableS} show the morphology of the simulated AP waves in time and space, respectively, comparing LD and NLD model at the four selected temperatures. As expected from the model formulation:
i) the shape of the AP wave does not change neither due to temperature nor to the nonlinear diffusion operator;
ii) temperature acts on both the temporal duration and the spatial distribution of the AP wave, in particular widening the AP duration at lower temperatures;
iii) NLD acts on the spatial properties of the AP wave and in particular on the repolarization phase of the wave~\cite{karma:1994};
iv) the combined effect of NLD and thermo-electric coupling entails a non-unique variation in space of the AP wave and produces higher differences at lower temperatures.

In order to emphasize the last observation, Fig.~\ref{fig:1c} shows a zoomed view of the back-end of the AP wave in space for a finer gradient of the temperatures: $T\in[31\div0.5\div40]^\circ$C. Linear and nonlinear diffusion are provided (dashed, solid) and two temperatures $33,34^\circ$C are highlighted. In particular, we observe that the linear model does not differentiate $33,33.5^\circ$C and $34,34.5^\circ$C thus resulting in the same AP wave length for any value of the intersecting threshold chosen (see Fig.~\ref{fig:1d}). In addition, this information is confirmed by the line integral of the normalized AP wave in space as shown in Fig.~\ref{fig:1e}.
It is worth noticing that the combined effect of nonlinear diffusion and thermo-electric coupling seems to regularize the non-monotone behavior observed for the linear model when a single propagating wave is analyzed. Richer information can be recovered when the system undergoes fast pacing or sustains rotating spirals.

\begin{figure*}[t!]
	\subfigure[]{\includegraphics[width=0.38\textwidth]{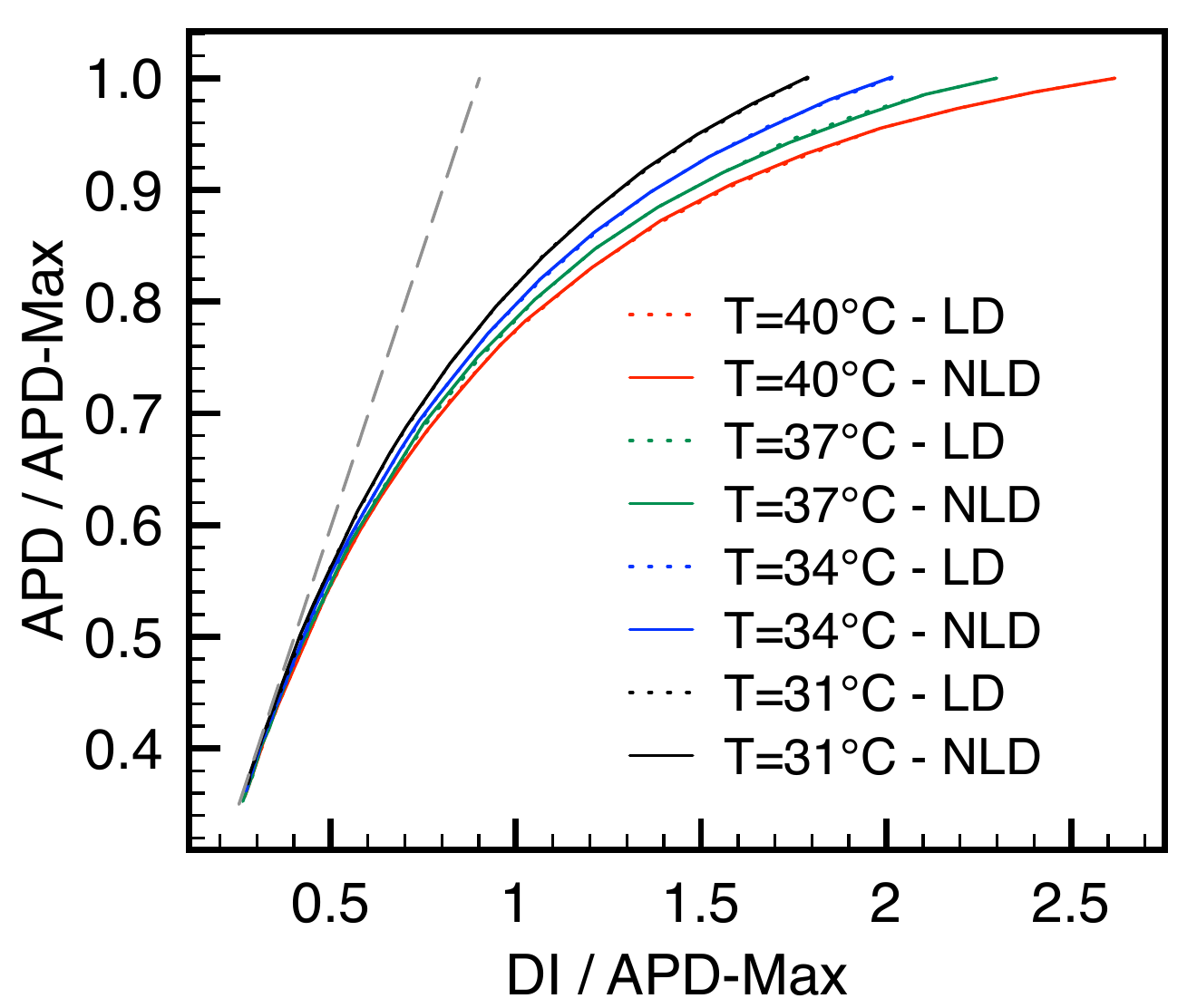}\label{fig:restAPDN}}
	\subfigure[]{\includegraphics[width=0.38\textwidth]{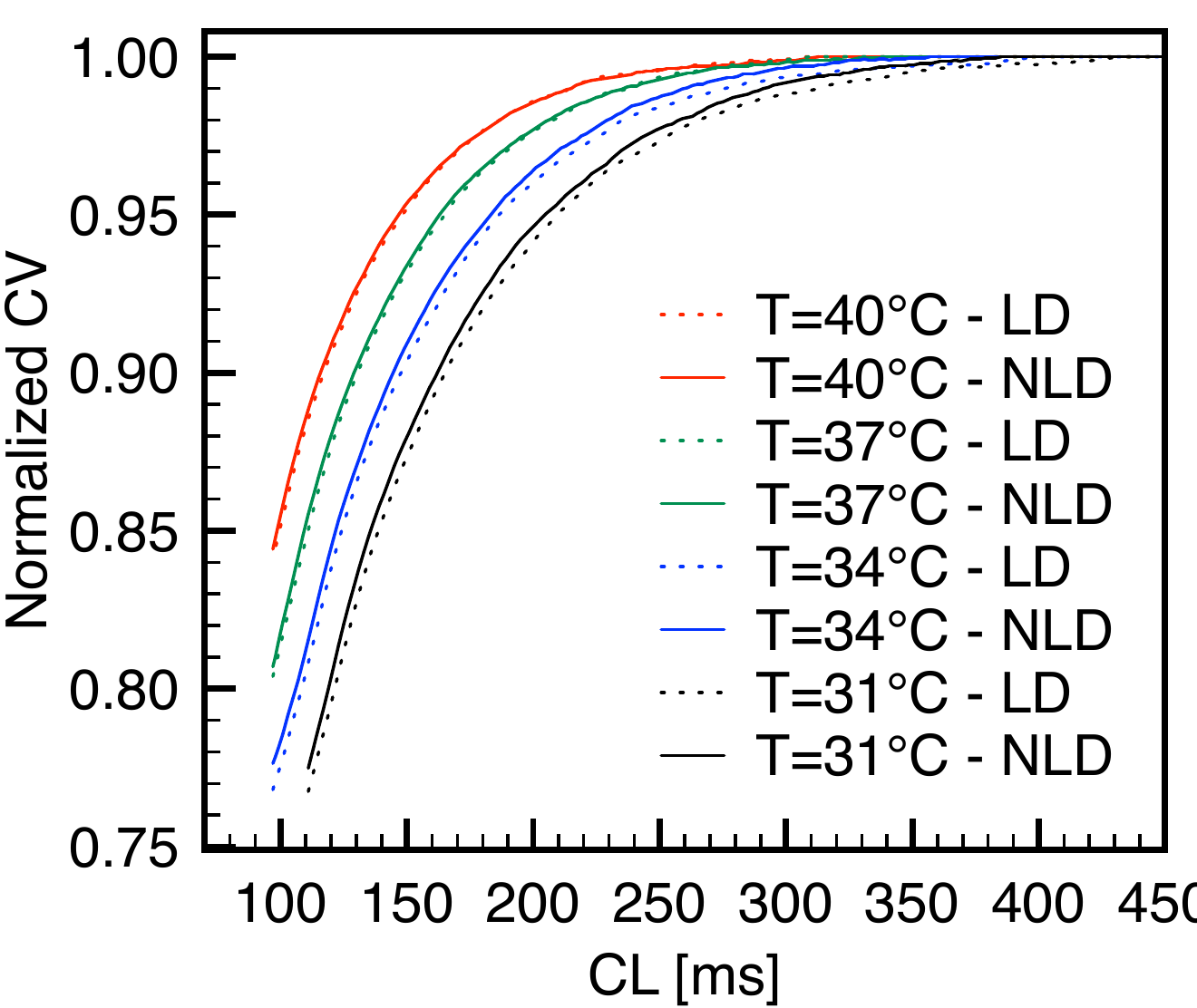}\label{fig:restCVN}}
	\subfigure{\includegraphics[width=1\textwidth]{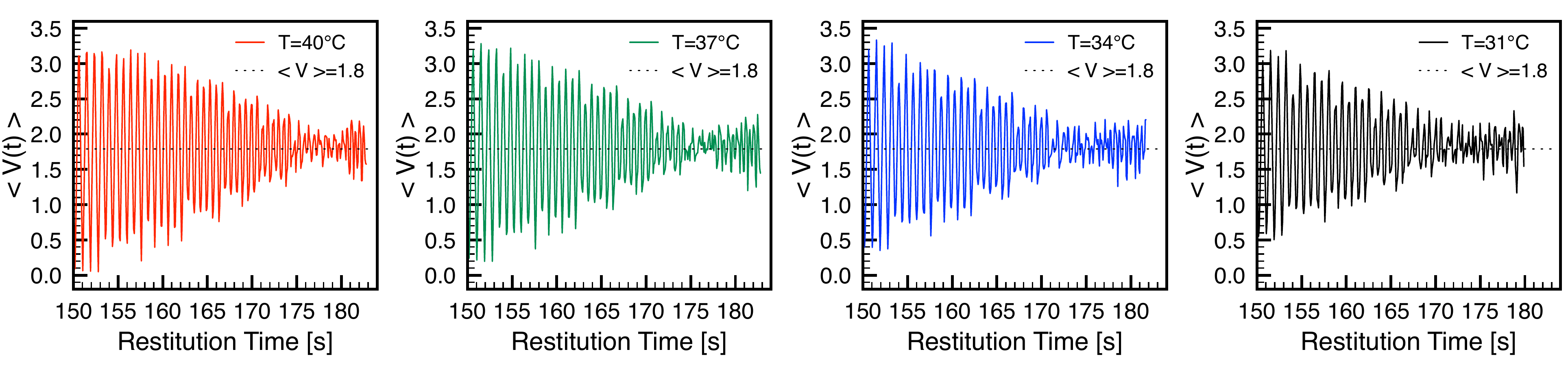}\label{fig:cableAPintNL}}
	\caption{\label{fig:rest} 
	Normalized restitutions curves obtained via a full pacing down stimulation protocol comparing linear and nonlinear diffusion at different temperatures.
	(a,b) APD-DI and CV-CL normalized maps with respect to the maximum value of APD (APD-Max) and of CV (CV-Max) for the four selected temperatures.
	In panel (a) the gray dashed line indicates slope 1.
	(c)
	Spatial average value of the excitation wave, $\Vm$, calculated during the full restitution protocol (with $n=5$ stimulations at constant CL) for the nonlinear model (the linear model behaves similarly). Each panel refers to a specific temperature value indicated in the legend.}
\end{figure*}

Figures~\ref{fig:restAPDN} and \ref{fig:restCVN} provide the normalized APD and CV restitution curves~\cite{karma:1994} obtained via the full pacing down protocol conducted over a 1D cable (APD and CV values are taken from the last activation wave). As expected from the point-wise temporal analysis shown in Fig.~\ref{fig:cableT}, LD and NLD do not induce significant differences in the APD restitution curves that usually fit experimental data~\cite{vinet:1990b}. Such a quantity, in fact, mainly relies on the front-end of the excitation wave which is known to be affected by nonlinear diffusion operators primarily on the foot of the AP wave spatial distribution~\cite{hurtado:2016,bueno-orovio:2015}. Since in our simplified case such a nonlinearity is moderate, the resulting APD restitution curves overlap.

On the contrary, small deviations are observed on the CV restitution curves thus entailing the eventual role of diffusion in the spatio-temporal features of the system in accordance with Fig.~\ref{fig:cableS}.
Temperature shifts the APD and CV curves: longer APDs and smaller CVs are in fact obtained at lower temperatures. A reduced minimum pacing CL  and a faster decay of the CV curve are also observed at lower temperatures.
It is important to note here that these small deviations may result in greatly different behavior for more complex models accounting for heterogeneities and anisotropies or when long run simulations are conducted.

\begin{figure*}[t!]
	{\includegraphics[width=0.95\textwidth]{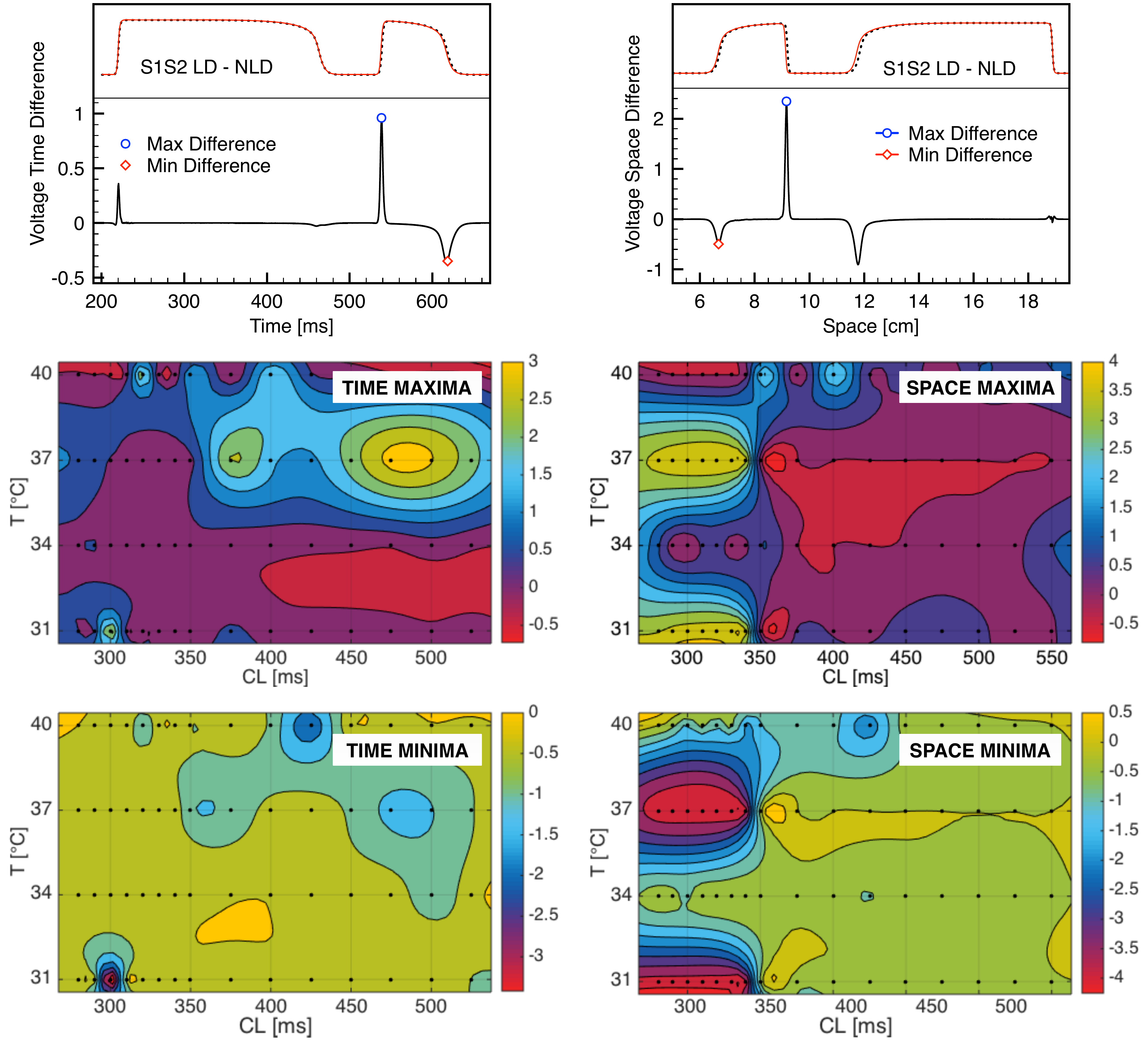}}
	\caption{\label{fig:PD3D}
	Interpolated contour plots (spline Matlab routine) of the temporal (left) and spatial (right) maxima and minima voltage phase differences computed on the S2 wave of the S1S2 restitution protocol (see top panel). Color code refers to the amplitude of the difference at the four selected temperatures. Simulated data are provided as black circles.}
\end{figure*}

The analysis of the restitution curves is here enriched with the dynamical calculation of the average action potential $\Vm$ on the simulated domain during periodic pacing. Figure~\ref{fig:cableAPintNL} highlights the last $30~s$ of the restitution protocol in terms of $\Vm$ (the sole nonlinear case is shown since similar to the linear case). It is worth noting that a stable oscillating signal is present before a reduction of the oscillations amplitude is reached, $\Vm\simeq1.8$, for all the selected temperature. This value corresponds to the physical condition for which the new excitation wave entering the domain from one boundary is balanced by the previous one leaving the domain from the opposite boundary. However, according to the restitution maps described before, the transition towards such a regime occurs at earlier times (e.g. at larger pacing CL) for lower temperatures. In addition, the last simulated times (shortest CLs) are characterized by an eventual increase of the oscillations just before loss of activation. At comparison with experimental evidences~\cite{gizzi:2013} and more sophisticated thermo-electric models~\cite{fenton:2013}, these fast pacing regimes are in fact characterized by a variety of alternating behaviors and bifurcation properties~\cite{fenton:2001} we do not capture in the present two-dimensional formulation.

\subsection{Phase Analysis}

We consider now the S1S2 restitution protocol and analyze temporal and spatial phase differences (PD) of the AP wave.
In detail, we synchronize the time course and spatial distribution of the AP signal on the S1 activation wave and calculate the algebraic difference for the sole S2 activation thus resulting in the plots shown in the top panel of Fig.~\ref{fig:PD3D}. We consider then the sole maxima and minima values resulting from these differences and plot them in the phase space (CL,T) by running the analysis over the four selected temperatures and pacing periods adopted during the stimulation protocol. The resulting diagrams indicate regions (basins) with high variations between LD and NLD models. In particular, the temporal PD only indicates maximum differences at high pacing CLs and normal temperature, while spatial PD shows a strong correlation with the major differences obtained at $37,31^\circ$C for short CLs. This second observation is in agreement with the restitution analysis discussed before thus implying a wide spectrum of novel properties to arise from the combination of nonlinear diffusion and nonlinear thermo-electricity. 

\begin{figure*}[t!]
	\subfigure[]{\includegraphics[width=1\textwidth]{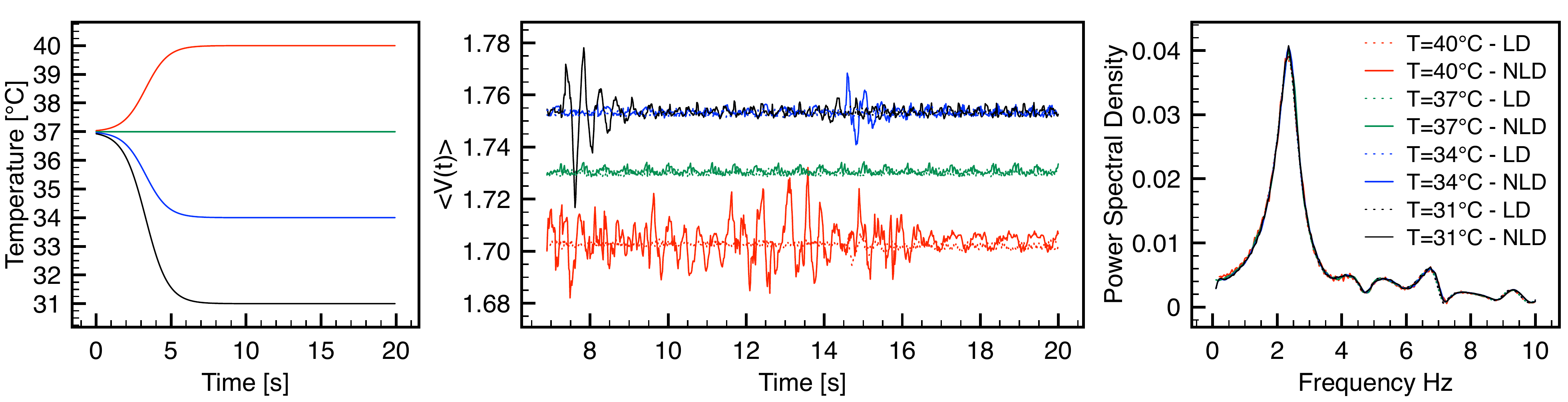}}
	\subfigure[]{\includegraphics[width=1\textwidth]{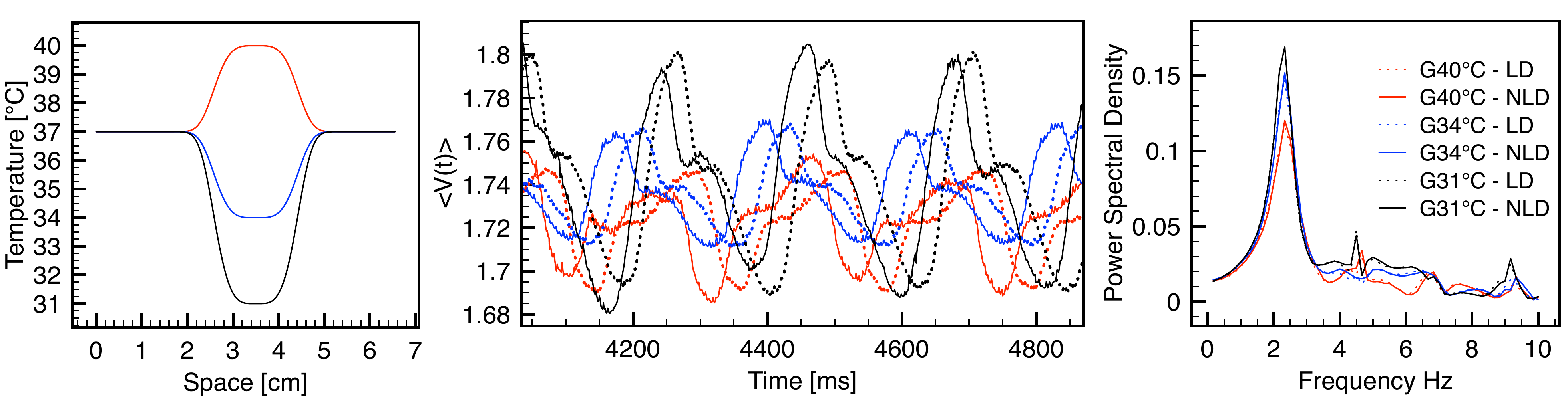}}
		\caption{\label{fig:exp} 
	Comparison of linear and nonlinear diffusion (dashed, solid line) on a ring of $L=6.55~cm$ according to the original Karma model~\cite{karma:1994}.
	(a) Transition from $T=37^\circ$C to hyper- and hypothermia static states (left);
	$\Vm$ time course of the system (color code as in previous plots) and corresponding power spectral density (right).
	(b) Static spatial distribution of temperatures for three difference scenarios (left). Corresponding time course of $\Vm$ and frequency spectrum (right).}
\end{figure*}

\begin{figure*}[t!]
	\includegraphics[width=1\textwidth]{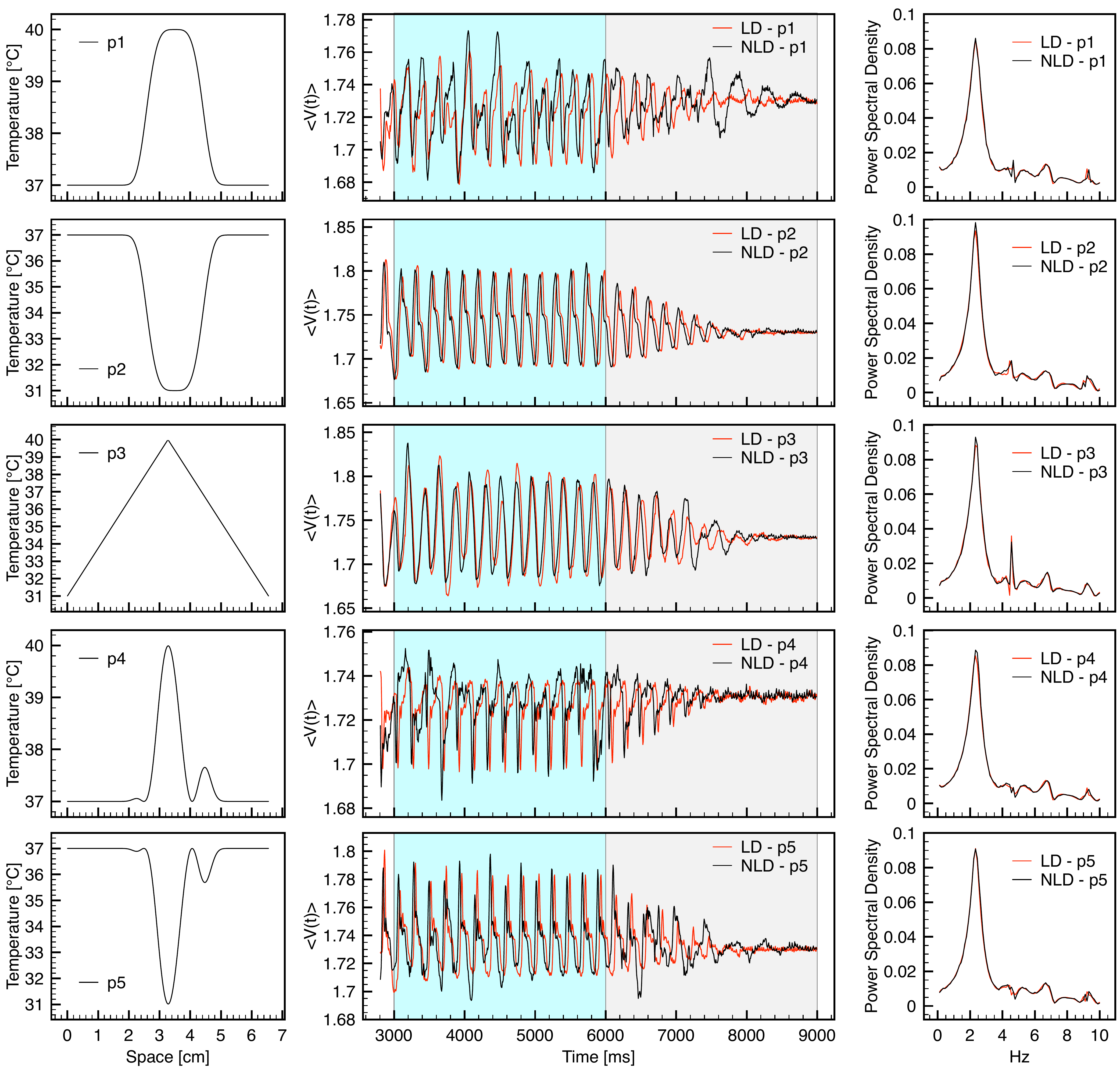}
	\caption{\label{fig:ringTtime} 
	Comparison of $\Vm$ (center) and corresponding spectra (right) for linear (LD) and nonlinear (NLD) diffusion models on a ring ($L=6.55~cm$) for different spatial thermal profiles (left) (p1-p5 stand for the selected thermal profile). A smooth dynamical transition from the reference physiological temperature, $37^\circ$C, is applied in $3~s$, kept for $3~s$ (light blue region) and finally removed in $3~s$ (gray region) such to recover the reference temperature. The plot reports the last $6~s$ of $\Vm$ in order to appreciate signal differences.}
\end{figure*}

\subsection{1D Ring Analysis}
As usual in the context of cardiac alternans and arrhythmias~\cite{winfree:1987}, in this section we investigate the dynamical properties of our thermo-electric system by analyzing a 1D cable with periodic boundary conditions (ring). In view of multi-dimensional analyses, we also consider different static and dynamic spatial distributions of the temperature.

Figure \ref{fig:exp} provides the time course of $\Vm$ and the corresponding frequency spectrum for steady state uniform and distributed temperatures (a,b). The analysis shows that the rotating wave stabilizes on different average values according to the temperature parameter  $T$. Persisting oscillations are present for the NLD case (solid) with respect to the LD one (dashed) and in particular at $40^\circ$C. However, the dominant frequencies of the system in the range $0\div10~Hz$ are captured in a similar manner and curves overlap (right panel).
However, if a spatial distribution of temperatures is considered in the ring (warmer and colder regions in panel b), then a clear shift between LD and NLD cases arises as well as the appearance of novel intermedia peaks in the corresponding frequency spectra. Of note an intermedia peak at about $5~Hz$ is present underlying an increased heterogeneity of the tissue affecting the characteristic time scales of the system during arrhythmias~\cite{karma:1994}.

An extended analysis of rotating waves in the 1D ring setting is further provided in Fig.~\ref{fig:ringTtime}. We compare different spatial distributions of the parameter $T$ with the additional investigation of thermal dynamical variation. We impose a smooth dynamical transition from the reference physiological temperature, $37^\circ$C, in $3~s$; we keep the new thermal state for $3~s$ (light blue region) and we finally recover the reference temperature in $3~s$ (gray region). Though longer time profiles may be investigated, the selected $3~s$ intervals allow the system to stabilize in the new rotating spiral configuration. It is worth noting here that we do not consider the effect of thermal diffusion and tissue perfusion which would require the additional coupling of a Pennes bio-equation~\cite{pennes,gizzi:2010}, but assume the thermal transition as instantaneous.

The results confirm the shift between LD and NLD models and indicate that the LD case is able to recover the original average value in shorter time than the NLD one. Such a behavior, with additional oscillations observed for the NLD  $\Vm$ signal, anticipates the arrhythmogenic propensity of the nonlinear diffusion model that we will highlight via two-dimensional simulations in the next section.

\subsection{Spiral Breakup}
\begin{figure*}[t!]
	\includegraphics[width=1\textwidth]{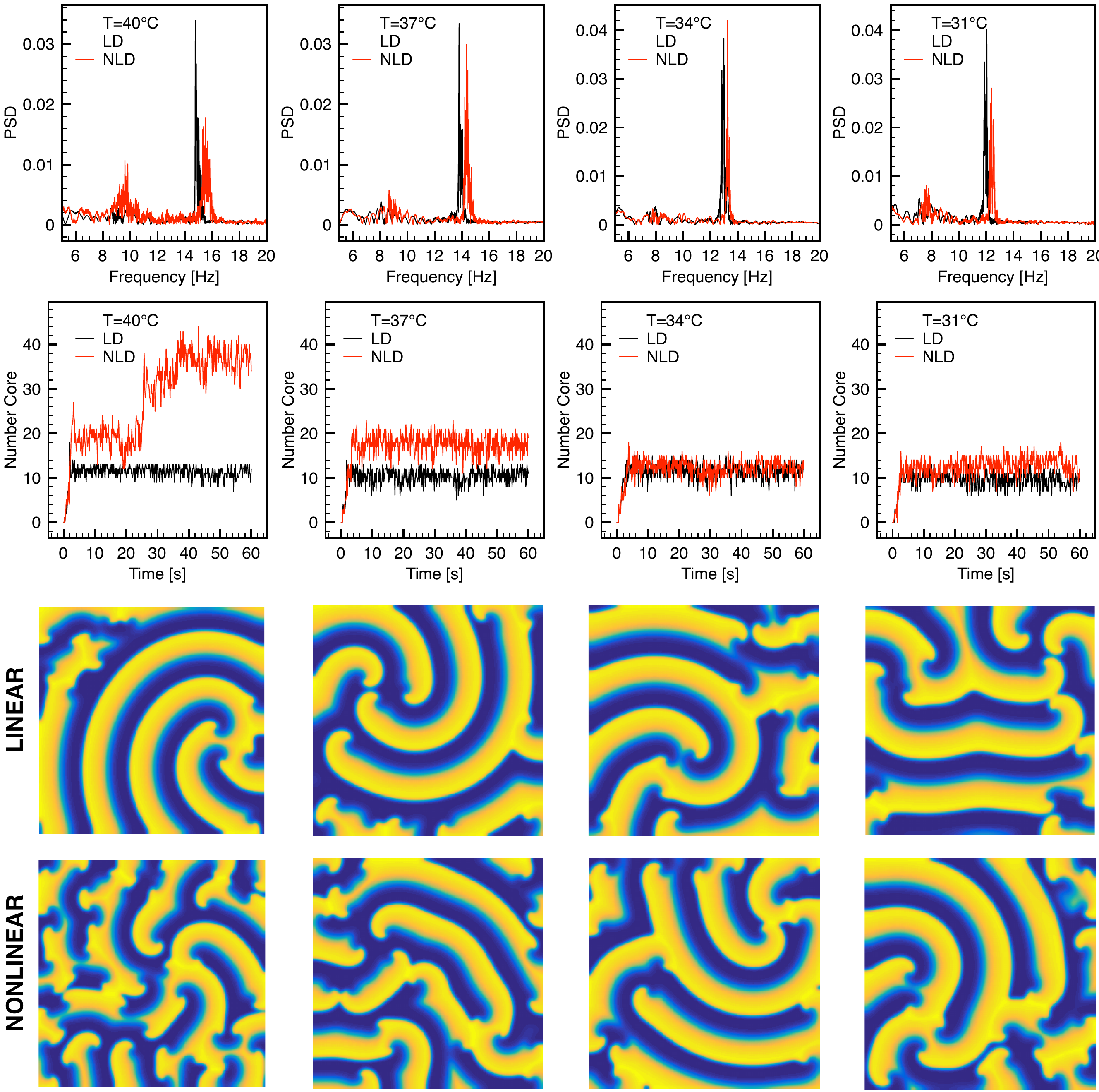}
	\caption{\label{fig:2Dspiral} 
	Long run 2D spiral breakup simulations ($6.5\times6.5~cm^2$) implemented in Fortran finite differences ($dt=0.1~ms$, $dx=0.025~cm$) comparing LD and NLD models at four selected temperatures.
	(Top) $\Vm$ power spectrum density (PSD) in the frequency range $5\div20~Hz$;
	(Center) numerosity of spiral cores;
	(Bottom) spatial distribution of the excitation waves at the final simulated time ($60~s$).}
\end{figure*}

We conclude the numerical investigation of the nonlinear dynamical system with long run simulations ($60~s$ of physical time) of sustained 2D spiral breakup at different temperatures. The results are shown in Fig.~\ref{fig:2Dspiral} in terms of $\Vm$ spectrum, spiral core numerosity and spatial distribution of the excitation waves. The LD and NLD models exhibit similar trends of $\Vm$ (not shown) but the resulting spectra are characterized by higher frequencies in the NLD case. In particular, both low ($\sim7\div9~Hz$) and high ($\sim13\div16~Hz$) peaks are consistent with similar analyses~\cite{karma:1994,filippi:2014} though a shifted dominant frequency of about $1~Hz$ is observed between linear and nonlinear case. This result generalizes the analyses conducted in the previous sections on the ring and is corroborated by the core numerosity plots highlighting a higher propensity of the system to produce  spiral breakup at higher temperatures since the wave lengths are shorter and the repolarization times are faster.
Accordingly, the spatial distribution of spiral waves shows different fibrillating scenarios between LD and NLD cases. 

It is important to observe here that the sole computed restitution curves are not able to predict these differences in the spatio-temporal dynamics. In fact, via the thermo-electric coupling we introduced an additional nonlinear effect in the system's temporal properties whereas a supplemental nonlinearity is enforced in the spatial flux via a dispersal model. The resulting dynamical system is therefore characterized by more complex emerging behaviors and additional nonlinearities and future studies are demanded in this direction on more physiological models.

\section{\label{sec:conclusions}Conclusions}

In summary, we have introduced a generalized two-variable model of action potential propagation accounting for thermo-electric coupling and nonlinear diffusion which reproduces characteristic restitution and dispersion properties of cardiac tissues. The simplified structure adopted allows to limit the parameter's space to the sole new elements ruling thermo-electricity within the framework of non-Fickian fluxes. However, several missing physiological factors are not captured and extended computational studies are envisioned with more physiological models~\cite{bueno:2008,fenton:2008}.

On this ground, we have investigated the dynamical origin of spiral wave breakup and compared long run simulations of linear and nonlinear diffusion under different thermal regimes. The main conclusion is that the onset of spiral wave breakup is enhanced at the highest and lowest temperatures tested by the nonlinear model. In addition, the resulting fibrillating scenario is affected by the non-Fickian flux in a non predictive way. By analogy with similar theoretical works~\cite{karma:1994,bueno-orovio:2014,bueno-orovio:2014a,bueno-orovio:2015,bueno-orovio:2016}, the present contribution suggest a direct connection between the emerging behavior at the tissue level with the dynamical response of single cells.

The second main conclusion of the work is that the linear and nonlinear dynamical systems adjust with slightly different behaviors when the temperature assumes heterogeneous distributions or is modulated in time.
It is worth noting that spatial heterogeneities and anisotropies may potentially play an important role in the present setting~\cite{fenton:1998,fenton:2002} and we expect to address this particular aspect in forthcoming contributions.

Of particular importance for understanding and control of cardiac arrhythmias, the study of cardiac alternans and their multiple transitions towards fibrillation would represent an additional important element to investigate within the context of thermo-electricity and nonlinear diffusion. In this perspective, three- or four-variable phenomenological models are the best candidates to explore the key features of such a complex dynamical system.


\begin{acknowledgments}
The authors thank Flavio H. Fenton (Georgia Institute of Technology, USA) and Elizabeth M. Cherry (Rochester Institute of Technology, USA) for helpful discussions and comments. This work was supported by funding from the Italian National Group of Mathematical Physics (GNFM-INdAM), the International Center for Relativistic Astrophysics Network (ICRANet) and the London Mathematical Society through its Grant Scheme 4 - Research in Pairs.
\end{acknowledgments}

\bibliography{chaos}

\begin{thebibliography}{58}
\providecommand{\natexlab}[1]{#1}
\providecommand{\url}[1]{\texttt{#1}}
\expandafter\ifx\csname urlstyle\endcsname\relax
  \providecommand{\doi}[1]{doi: #1}\else
  \providecommand{\doi}{doi: \begingroup \urlstyle{rm}\Url}\fi

\bibitem[Bendahmane and Karlsen(2006)]{bendahmane:2006}
M.~Bendahmane and K.~H. Karlsen.
\newblock Analysis of a class of degenerate reaction-diffusion systems and the
  bidomain model of cardiac tissue.
\newblock \emph{Networks and Heterogeneous Media}, 1:\penalty0 185--218, 2006.

\bibitem[Bendahmane et~al.(2010)Bendahmane, B\"{u}rger, and
  Ruiz-Baier]{ruiz:2010}
M.~Bendahmane, R.~B\"{u}rger, and R.~Ruiz-Baier.
\newblock A multiresolution space-time adaptive scheme for the bidomain model
  in electrocardiology.
\newblock \emph{Numerical Methods for Partial Differential Equations},
  26:\penalty0 1377--1404, 2010.

\bibitem[Bigelow et~al.(1950)Bigelow, Callaghan, and Hopps]{bigelow:1950}
W.~G. Bigelow, J.~C. Callaghan, and J.~A. Hopps.
\newblock General hypothermia for experimental intracardiac surgery; the use of
  electrophrenic respirations, an artificial pacemaker for cardiac standstill
  and radio-frequency rewarming in general hypothermia.
\newblock \emph{Ann Surg}, 132:\penalty0 531--539, 1950.

\bibitem[Biktashev et~al.(1999)Biktashev, Biktasheva, Holden, Tsyganov,
  Brindley, and Hill]{biktashev:1999}
V.~N. Biktashev, I.~V. Biktasheva, A.~V. Holden, M.~A. Tsyganov, J.~Brindley,
  and N.~A. Hill.
\newblock Spatiotemporal irregularity in an excitable medium with shear flow.
\newblock \emph{Physical Review E}, 60:\penalty0 1987, 1999.

\bibitem[Biktasheva(2000)]{biktasheva:2000}
I.~V. Biktasheva.
\newblock Drift of spiral waves in the complex ginzburg-landau equation due to
  media inhomogeneities.
\newblock \emph{Physical Review E}, 62:\penalty0 8800, 2000.

\bibitem[Biktasheva et~al.(2015)Biktasheva, Dierckx, and
  Biktashev]{biktasheva:2015}
I.~V. Biktasheva, H.~Dierckx, and V.~N. Biktashev.
\newblock Drift of spiral waves in the complex ginzburg-landau equation due to
  media inhomogeneities.
\newblock \emph{Physical Review Letters}, 114:\penalty0 068302, 2015.

\bibitem[Bueno-Orovio et~al.(2008)Bueno-Orovio, Cherry, and Fenton]{bueno:2008}
A.~Bueno-Orovio, E.~M. Cherry, and F.~H. Fenton.
\newblock Minimal model for human ventricular action potentials in tissue.
\newblock \emph{Journal of Theoretical Biology}, 7:\penalty0 544--560, 2008.

\bibitem[Bueno-Orovio et~al.(2014{\natexlab{a}})Bueno-Orovio, Kay, and
  Burrage]{bueno-orovio:2014a}
A.~Bueno-Orovio, D.~Kay, and K.~Burrage.
\newblock Fourier spectral methods for fractional-in-space reaction-diffusion
  equations.
\newblock \emph{BIT Numerical Mathematics}, 54:\penalty0 937--954,
  2014{\natexlab{a}}.

\bibitem[Bueno-Orovio et~al.(2014{\natexlab{b}})Bueno-Orovio, Kay, Grau,
  Rodriquez, and Burrage]{bueno-orovio:2014}
A.~Bueno-Orovio, D.~Kay, V.~Grau, B.~Rodriquez, and K.~Burrage.
\newblock Fractional diffusion models of cardiac electrical propagation: role
  of structural heterogeneity in dispersion of repolarization.
\newblock \emph{Journal of the Royal Society Interface}, 11:\penalty0 20140352,
  2014{\natexlab{b}}.

\bibitem[Bueno-Orovio et~al.(2016)Bueno-Orovio, Teh, Schneider, Burrage, and
  Grau]{bueno-orovio:2016}
A.~Bueno-Orovio, I.~Teh, J.~E. Schneider, K.~Burrage, and V.~Grau.
\newblock Anomalous diffusion in cardiac tissue as an index of myocardial
  microstructure.
\newblock \emph{IEEE Transactions Medical Imaging}, 35:\penalty0 2200--2207,
  2016.

\bibitem[Chen et~al.(In Press)Chen, Gray, Uzelac, Herndon, and
  Fenton]{chen:2017}
D.~D. Chen, R.~A. Gray, I.~Uzelac, C.~Herndon, and F.~H. Fenton.
\newblock Mechanism for amplitude alternans in electrocardiograms and the
  initiation of spatiotemporal chaos.
\newblock \emph{Physical Review Letters}, In Press.

\bibitem[Cherry and Fenton(2007)]{cherry:2007}
E.~M. Cherry and F.~H. Fenton.
\newblock A tale of two dogs: analyzing two models of canine ventricular
  electrophysiology.
\newblock \emph{American Journal of Physiology-Heart and Circulatory
  Physiology}, 292:\penalty0 H43--H55, 2007.

\bibitem[Cherubini et~al.(2012)Cherubini, Filippi, and Gizzi]{cherubini:2012}
C.~Cherubini, S.~Filippi, and A.~Gizzi.
\newblock Electroelastic unpinning of rotating vortices in biological excitable
  media.
\newblock \emph{{Physical Review E}}, 85:\penalty0 031915, 2012.

\bibitem[Clayton et~al.(2011)Clayton, Bernus, Cherry, Dierckx, Fenton,
  Mirabella, Panfilov, Sachse, Seemann, and Zhang]{clayton:2011}
R.~H. Clayton, O.~Bernus, E.~M. Cherry, H.~Dierckx, F.~H. Fenton, L.~Mirabella,
  A.~V. Panfilov, F.~B. Sachse, G.~Seemann, and H.~Zhang.
\newblock Models of cardiac tissue electrophysiology: Progress, challenges and
  open questions.
\newblock \emph{Progress in Biophysics and Molecular Biology}, 104:\penalty0
  22--48, 2011.

\bibitem[Cusimano et~al.(2015)Cusimano, Bueno-Orovio, Turner, and
  Burrage]{bueno-orovio:2015}
N.~Cusimano, A.~Bueno-Orovio, I.~Turner, and K.~Burrage.
\newblock On the order of the fractional laplacian in determining the
  spatio-temporal evolution of a space-fractional model of cardiac
  electrophysiology.
\newblock \emph{PLoS ONE}, 10:\penalty0 e0143938, 2015.

\bibitem[Dehin(2006)]{dehin:2006}
S.~Dehin, editor.
\newblock \emph{Cardiovascular Gap Junctions}, volume~42 of \emph{Advances in
  Cardiology}.
\newblock Karger, 2006.

\bibitem[Deshpande et~al.(2006)Deshpande, McMeeking, and Evans]{deshpande:2006}
V.~S. Deshpande, R.~M. McMeeking, and A.~G. Evans.
\newblock A bio-chemo-mechanical model for cell contractility.
\newblock \emph{PNAS}, 103:\penalty0 14015--14020, 2006.

\bibitem[Deshpande et~al.(2008)Deshpande, Mrksich, McMeeking, and
  Evans]{deshpande:2008}
V.~S. Deshpande, M.~Mrksich, R.~M. McMeeking, and A.~G. Evans.
\newblock A bio-mechanical model for coupling cell contractility with focal
  adhesion formation.
\newblock \emph{Journal of the Mechanics and Physics of Solids}, 56:\penalty0
  1484--1510, 2008.

\bibitem[Dupraz and Jacquemet(2014)]{dupraz:2014}
M.~Dupraz and V.~Jacquemet.
\newblock Geometrical measurement of cardiac wavelength in reaction-diffusion
  models.
\newblock \emph{Chaos}, 24:\penalty0 033133, 2014.

\bibitem[Fenton and Cherry(2008)]{fenton:2008}
F.~H. Fenton and E.~M. Cherry.
\newblock Models of cardiac cell.
\newblock \emph{Scholarpedia}, 3:\penalty0 1868, 2008.

\bibitem[Fenton and Karma(1998)]{fenton:1998}
F.~H. Fenton and A.~Karma.
\newblock Vortex dynamics in three-dimensional continuous myocardium with fiber
  rotation: Filament instability and fibrillation.
\newblock \emph{{Chaos}}, 8:\penalty0 1054, 1998.

\bibitem[Fenton et~al.(2002)Fenton, Cherry, Hastings, and Evans]{fenton:2002}
F.~H. Fenton, E.~M. Cherry, H.~M. Hastings, and S.~G. Evans.
\newblock Multiple mechanisms of spiral wave breakup in a model of cardiac
  electrical activity.
\newblock \emph{Chaos: An Interdisciplinary Journal of Nonlinear Science},
  12:\penalty0 852--892, 2002.

\bibitem[Fenton et~al.(2005)Fenton, Cherry, Karma, and Rappel]{fenton:2005}
F.~H. Fenton, E.~M. Cherry, A.~Karma, and W.~J. Rappel.
\newblock Modeling wave propagation in realistic heart geometries using the
  phase-field method.
\newblock \emph{Chaos: An Interdisciplinary Journal of Nonlinear Science},
  15:\penalty0 013502, 2005.

\bibitem[Fenton et~al.(2013)Fenton, Gizzi, Cherubini, Pomella, and
  Filippi]{fenton:2013}
F.~H. Fenton, A.~Gizzi, C.~Cherubini, N.~Pomella, and S.~Filippi.
\newblock Role of temperature on nonlinear cardiac dynamics.
\newblock \emph{Physical Review E}, 87:\penalty0 042709, 2013.

\bibitem[Filippi et~al.(2014)Filippi, Gizzi, Cherubini, Luther, and
  Fenton]{filippi:2014}
S.~Filippi, A.~Gizzi, C.~Cherubini, S.~Luther, and F.~H. Fenton.
\newblock Mechanistic insights into hypothermic ventricular fibrillation: The
  role of temperature and tissue size.
\newblock \emph{Europace}, 16:\penalty0 424--434, 2014.

\bibitem[Gizzi et~al.(2010)Gizzi, Cherubini, Migliori, Alloni, Portuesi, and
  Filippi]{gizzi:2010}
A.~Gizzi, C.~Cherubini, S.~Migliori, R.~Alloni, R.~Portuesi, and S.~Filippi.
\newblock On the electrical intestine turbulence induced by temperature
  changes.
\newblock \emph{{Physical Byology}}, 7:\penalty0 016011, 2010.

\bibitem[Gizzi et~al.(2013)Gizzi, Cherry, Jr, Luther, Filippi, and
  Fenton]{gizzi:2013}
A.~Gizzi, E.~M. Cherry, Gilmour R.~F. Jr, S.~Luther, S.~Filippi, and F.~H.
  Fenton.
\newblock Effects of pacing site and stimulation history on alternans dynamics
  and the development of complex spatiotemporal patterns in cardiac tissue.
\newblock \emph{Frontiers in Physiology}, 4:\penalty0 71, 2013.

\bibitem[Group(1997)]{lancet:1997}
The~Eurowinter Group.
\newblock Cold exposure and winter mortality from ischaemic heart disease,
  cerebrovascular disease, respiratory disease, and all causes in warm and cold
  regions of europe.
\newblock \emph{Lancet}, 349:\penalty0 1341--1346, 1997.

\bibitem[Hurtado et~al.(2016)Hurtado, Castro, and Gizzi]{hurtado:2016}
D.~E. Hurtado, S.~Castro, and A.~Gizzi.
\newblock Computational modeling of non-linear diffusion in cardiac
  electrophysiology: A novel porous-medium approach.
\newblock \emph{Computer Methods in Applied Mechanics and Engineering},
  300:\penalty0 70--83, 2016.

\bibitem[Karma(1991)]{karma:1991}
A.~Karma.
\newblock Universal limit of spiral wave propagation in excitable media.
\newblock \emph{Physical Review Letters}, 66:\penalty0 2274--2277, 1991.

\bibitem[Karma(1992)]{karma:1992}
A.~Karma.
\newblock Scaling regime of spiral wave propagation in single.diffusive media.
\newblock \emph{Physical Review Letters}, 68:\penalty0 397--400, 1992.

\bibitem[Karma(1993)]{karma:1993}
A.~Karma.
\newblock Spiral breakup in model equations of action potential propagation in
  cardiac tissue.
\newblock \emph{Physical Reviews Letters}, 71:\penalty0 1103, 1993.

\bibitem[Karma(1994)]{karma:1994}
A.~Karma.
\newblock Electrical alternans and spiral wave breackup in cardiac tissue.
\newblock \emph{Chaos: An Interdisciplinary Journal of Nonlinear Science},
  4:\penalty0 461--472, 1994.

\bibitem[Karma et~al.(1994)Karma, Levine, and Zou]{karma:1994D}
A.~Karma, H.~Levine, and X.~Zou.
\newblock Theory of pulse instabilities in electrophysiological models of
  excitable tissues.
\newblock \emph{Physica D}, 73:\penalty0 113--127, 1994.

\bibitem[Keener and Sneyd(2009)]{keener}
J.~Keener and J.~Sneyd.
\newblock \emph{Mathematical Physiology}.
\newblock Spinger--Verlag, 2009.

\bibitem[Krinsky and Pumir(1998)]{pumir:1998}
V.~Krinsky and A.~Pumir.
\newblock Models of defibrillation of cardiac tissue.
\newblock \emph{Chaos}, 8:\penalty0 188--203, 1998.

\bibitem[Luther et~al.(2011)Luther, Fenton, Kornreich, Squires, Bitthn,
  Hornung, Zabel, Flanders, Gladuli, Campoy, Cherry, Luther, Hasenfuss,
  Krinsky, Pumir, Jr, and Bodenschatz]{luther:2011}
S.~Luther, F.~H. Fenton, B.~G. Kornreich, A.~Squires, P.~Bitthn, D.~Hornung,
  M.~Zabel, J.~Flanders, A.~Gladuli, L.~Campoy, E.~M. Cherry, G.~Luther,
  G.~Hasenfuss, V.~I. Krinsky, A.~Pumir, Gilmour R.~F. Jr, and Bodenschatz.
\newblock Low-energy control of electrical turbulence in the heart.
\newblock \emph{{Nature Letter}}, 475\penalty0 (235--239):\penalty0 1000--1007,
  2011.

\bibitem[Murray(2002)]{murray}
J.~D. Murray.
\newblock \emph{Mathematical Biology}.
\newblock Springer International Publishing, 2002.

\bibitem[Niederer and et~al.(2011)]{niederer:2011}
A.~A. Niederer and et~al.
\newblock Verification of cardiac tissue electrophysiology simulators using an
  n-version benchmark.
\newblock \emph{Phil. Trans. R. Soc. A}, 369:\penalty0 4331--4351, 2011.

\bibitem[Pastore and Rosenbaum(2000)]{pastore:2000}
J.~M. Pastore and D.~S. Rosenbaum.
\newblock Role of structural barriers in the mechanism of alternans- induced
  reentry.
\newblock \emph{{Circulation Research}}, 87:\penalty0 1157--1163, 2000.

\bibitem[Pastore et~al.(1999)Pastore, Girouard, Lautira, Akar, and
  Rosenbaum]{pastore:1999}
J.~M. Pastore, S.~D. Girouard, K.~R. Lautira, F.~G. Akar, and D.~S. Rosenbaum.
\newblock Mechanism linking t-wave alternans to the genesis of cardiac
  fibrillation.
\newblock \emph{Circulation}, 99:\penalty0 1385--1394, 1999.

\bibitem[Pennes(1948)]{pennes}
H.~H. Pennes.
\newblock Analysis of tissue and arterial blood temperatures in the resting
  human forearm.
\newblock \emph{J. Appl. Physiol.}, 1:\penalty0 93--122, 1948.

\bibitem[Potse et~al.(2006)Potse, Dub\'e, Richer, Vinet, and
  Gulrajani]{potse:2006}
M.~Potse, B.~Dub\'e, J.~Richer, A.~Vinet, and R.~M. Gulrajani.
\newblock A comparison of monodomain and bidomain reaction-diffusion models for
  action potential propagation in the human heart.
\newblock \emph{IEEE Trans. Biomed. Engineering}, 53:\penalty0 2425--2435,
  2006.

\bibitem[Potse et~al.(2007)Potse, Coronel, Falcao, LeBlanc, and
  Vinet]{potse:2007}
M.~Potse, R.~Coronel, S.~Falcao, A.-R. LeBlanc, and A.~Vinet.
\newblock The effect of lesion size and tissue remodeling on st deviation in
  partial-thickness ischemia.
\newblock \emph{Heart Rhythm}, 4:\penalty0 200--206, 2007.

\bibitem[Pullan et~al.(20015)Pullan, Buist, and Cheng]{pullan}
A.~J. Pullan, M.~L. Buist, and L.~K. Cheng.
\newblock \emph{Mathematically Modelling the Electrical Activity of the Heart:
  From Cell to Body Surface and Back Again}.
\newblock World Scientific, 20015.

\bibitem[Pumir et~al.(2005)Pumir, Arutunyan, Krinsky, and
  Sarvazyan]{pumir:2005}
A.~Pumir, A.~Arutunyan, V.~Krinsky, and N.~Sarvazyan.
\newblock Genesis of ectopic waves: Role of coupling, automaticity, and
  heterogeneity.
\newblock \emph{Biophysical Journal}, 89:\penalty0 2332--2349, 2005.

\bibitem[Pumir et~al.(2010)Pumir, Sinha, Sridhar, Argentina, H{\"o}rning,
  Filippi, Cherubini, Luther, and Krinsky]{pumir:2010}
A.~Pumir, S.~Sinha, S.~Sridhar, M.~Argentina, M.~H{\"o}rning, S.~Filippi,
  C.~Cherubini, S.~Luther, and V.I. Krinsky.
\newblock Wave-train-induced termination of weakly anchored vortices in
  excitable media.
\newblock \emph{{Physical Review E}}, 81:\penalty0 010901, 2010.

\bibitem[Qu et~al.(2014)Qu, Hu, Garfinkel, and Weiss]{qu:2014}
Z.~Qu, G.~Hu, A.~Garfinkel, and J.~N. Weiss.
\newblock Nonlinear and stochastic dynamics in the heart.
\newblock \emph{Physics Reports}, 543:\penalty0 61--162, 2014.

\bibitem[Quail et~al.(2014)Quail, Shrier, and Glass]{quail:2014}
T.~Quail, A.~Shrier, and L~Glass.
\newblock Spatial symmetry breaking determines spiral wave chirality.
\newblock \emph{Physical Review Letters}, 113:\penalty0 158101, 2014.

\bibitem[Reuler(1978)]{reuler:1978}
J.B. Reuler.
\newblock Hypothermia: pathophysiology, clinical settings, and management.
\newblock \emph{Annals of Internal Medicine}, 89:\penalty0 519--527, 1978.

\bibitem[Rohr(2004)]{rohr:2004}
S.~Rohr.
\newblock Role of gap junctions in the propagation of the cardiac action
  potential.
\newblock \emph{Cardiovascular Research}, 62:\penalty0 309--322, 2004.

\bibitem[S\'aez and Khul(2016)]{saez:2016}
P.~S\'aez and E.~Khul.
\newblock Computational modeling of acute myocardial infarction.
\newblock \emph{Computer Methods in Biomechanics and Biomedical Engineering},
  19:\penalty0 1107--1115, 2016.

\bibitem[Severs et~al.(2008)Severs, Bruce, Dupont, and Rothery]{severs:2008}
N.~J. Severs, A.~F. Bruce, E.~Dupont, and S.~Rothery.
\newblock Remodelling of gap junctions and connexin expression in diseased
  myocardium.
\newblock \emph{Cardiovascular Research}, 80:\penalty0 9--19, 2008.

\bibitem[Takagi et~al.(2004)Takagi, Pumir, Paz\'o, Efimov, Nikolski, and
  Krinsky]{pumir:2004}
S.~Takagi, A.~Pumir, D.~Paz\'o, I.~Efimov, V.~Nikolski, and V.~Krinsky.
\newblock Unpinning and removal of a rotating wave in cardiac muscle.
\newblock \emph{Physical Review Letters}, 93:\penalty0 058101, 2004.

\bibitem[Vinet et~al.(1990{\natexlab{a}})Vinet, Chialvo, and
  Jalife]{vinet:1990}
A.~Vinet, D.~R. Chialvo, and J.~Jalife.
\newblock Irregular dynamics of excitation in biologic and mathematical models
  of cardiac cells.
\newblock \emph{Annals of the New York Academy of Sciences}, 601:\penalty0
  281--298, 1990{\natexlab{a}}.

\bibitem[Vinet et~al.(1990{\natexlab{b}})Vinet, Chialvo, Michaels, and
  Jalife]{vinet:1990b}
A.~Vinet, D.~R. Chialvo, D.~C. Michaels, and J.~Jalife.
\newblock Nonlinear dynamics of rate-dependent activation in models of single
  cardiac cells.
\newblock \emph{{Circulation Research}}, 67:\penalty0 1510--1524,
  1990{\natexlab{b}}.

\bibitem[Watanabe et~al.(2001)Watanabe, Fenton, Evans, and Karma]{fenton:2001}
M.~A. Watanabe, F.~H. Fenton, H.~M. Evans, S.~J.~Hastings, and A.~Karma.
\newblock Mechanisms for discordant alternans.
\newblock \emph{{Journal of Cardiovascular Electrophysiology}}, 12:\penalty0
  196--206, 2001.

\bibitem[Winfree(1987)]{winfree:1987}
A.~T. Winfree.
\newblock \emph{When time breaks down}.
\newblock Princeton University Press, 1987.

\end{thebibliography}

\end{document}